\documentstyle[twocolumn,epsf,seceq]{jpsj}

\def\runtitle{Numerical Study on Aging Dynamics in  the 3D Ising
  Spin-Glass Model. I.}
\def\runauthor{Tatsuo {\sc Komori}, Hajime {\sc Yoshino} and Hajime 
{\sc Takayama}}

\title{Numerical Study on Aging Dynamics in  the 3D Ising Spin-Glass
       Model. I. Energy Relaxation and Domain
       Coarsening}
\author{Tatsuo {\sc Komori}\footnote{Present address: Hydrographic
      Department, Maritime Safety Agency, 5-3-1 Tsukiji, Chuo-ku,
      Tokyo 104-0045},
 Hajime {\sc Yoshino} and Hajime 
{\sc Takayama}\footnote{E-mail: takayama@issp.u-tokyo.ac.jp}}
\inst{Institute for Solid State Physics, the University of Tokyo,
7-22-1 Roppongi, Minato-ku, Tokyo 106-8666}
\recdate{\today}
\abst{Using Monte Carlo simulations, we have studied the relaxation of 
energy of the three-dimensional Ising spin-glass model in aging process.
Our finite-size-scaling analysis on the isothermal energy decay after
the quench suggests strongly that the energy relaxes by coarsening of
domain walls in agreement with the droplet theory.
We have also evaluated relaxation times required for spin 
configurations in small systems, which are regarded as isolated
droplets, to flip globally. The results support the fundamental
assumption of the droplet theory that coarsening of 
domain walls, in case of spin glasses, is driven by
successive nucleation of thermally activated droplets.
}
\kword{spin glass, slow dynamics, aging, droplet theory}

\begin{document}
\sloppy
\maketitle

\section{Introduction}\label{sec:fss-1}

In recent years off-equilibrium dynamics, particularly aging dynamics, 
of spin glasses has attracted much attention theoretically,\cite{BCKM} 
numerically,\cite{MPR} and experimentally.\cite{EXP,Nordblad,Weissman}
Many experiments have confirmed that the response to small dc/ac 
magnetic field after quenching the system below the spin-glass 
transition temperature $T_{\rm c}$ shows aging effects persistently 
at least up to the largest possible time scale available in 
laboratories.
On the side of theories, one of the most notable advances  is the 
analytical theory\cite{BCKM,CK,FM} based on solvable mean-field 
spin-glass models, which has provided
non-trivial predictions on aging effects. 

Presumably one expects that the slow relaxation is caused by some
underlying collective spin excitations in the system.  
In the case of ordering processes in non-random systems like the
conventional Ising ferromagnet, the relaxational dynamics after 
quenching the system below the phase transition temperature is related 
to coarsening of domain walls which separate ordered domains of two 
different pure states.\cite{Bray-94}
The typical size of the domains, $R(t)$, at time  $t$ usually grows by
a certain power law. This behavior can be directly 
studied, for instance by scattering experiments.
The intriguing question is if such a picture of the coarsening of 
domain walls also holds for aging processes in spin glasses. 
Obviously, the above mentioned mean-field theory, which is exact only
in infinitely large dimension, is not helpful in this respect. 
The domain picture based on the so-called droplet
theory,\cite{BM-84,FH-88-EQ,FH-88-NE,Huse-91}
has provided an interesting phenomenology.
However the pure states of spin glasses cannot be found in practice
which makes it difficult to rationalize the domain picture in a direct way.

In the present study,~\cite{Komori-D} 
we mainly focus on the energy relaxation in 
the 3-dimensional EA spin-glass model during the aging process
simulated by Monte Carlo dynamics. 
The latter starts from arbitrary initial configurations at temperature 
$T$ below $T_{\rm c}$ which simulates
instantaneous quench to $T$ from $T=\infty$.
We have found that the relaxation of energy can be explained  
consistently in terms of coarsening of domain walls as in the case of usual 
phase ordering processes in non-random systems. 
Let us briefly describe the idea here.
Suppose a system is aged by elapsed time $t$ after the
quench, and a  typical size of domain walls has grown to $R(t)$. 
In the presence of domain walls, 
the energy per spin, $e_{T}(t)$, is larger than the equilibrium energy 
 $e_{T}^{(\infty)}$ at temperature $T$ by an amount
\begin{equation}
   \delta e_{T}(t) \equiv  e_{T}(t)-e_{T}^{(\infty)}
\propto \frac{\Upsilon{}(T)(R(t)/l_{0})^{\theta}}{(R(t)/l_{0})^{d}}.
\label{eq:fss-1-1}
\end{equation}
with $l_0$ being a certain microscopic unit of length.
Here the numerator on the right-hand side is the domain wall energy where 
$\Upsilon{}(T)$ is the stiffness constant and $\theta$ is the
characteristic exponent.  The denominator is the volume of a domain with
$d$ being the dimension of the space. 

We have studied systematically the energy per spin on relatively small 
system sizes $L$ and at relatively higher temperatures in the spin-glass 
phase. This enables us to see the behaviors of two time regimes within
our computational time window; the size-independent behavior 
of domain growth in the  dynamic regime $R(t) < L$ 
and the size-dependent equilibrium behavior in the static regime $L < R(t)$. 
It has turned out that the simulated data including in the crossover
region between the two regimes can be concisely
described by a finite-size-scaling (FSS) function. 
By multiplying eq.~(\ref{eq:fss-1-1}) by a scaling
function $\tilde{\delta e}(R(t)/L)$, the scaling ansatz is proposed as
\begin{equation}
  \delta e_{T}(t) = \Upsilon(T)
  \left(\frac{R(t)}{l_0}\right)^{\theta -d}
    \tilde{\delta e}\left(
  \frac{R(t)}{L}\right),
\label{eq:fss-tmp}
\end{equation}
with $R(t)=t^{1/z(T)}$.  The FSS analysis 
allows us to obtain the characteristic exponent $\theta$ for 
the domain wall energy. The value of $\theta$ turns out to be nearly
equal to the result by Bray and Moore 
at $T=0$.\cite{BM-84} These results strongly support that the aging process 
in the present spin-glass model is described well in terms of coarsening 
of domain walls, i.e., the droplet picture.

As described just above the FSS analysis on the energy relaxation
yields the power-law growth in $t$ of the typical domain size $R(t)$
with the exponent $z(T)$, which is consistent with that obtained by the 
previous numerical studies using 
replica-overlap.\cite{Huse-91,Kisker-96,Marinari-98-VFDT} 
We have also performed analysis on the replica-overlap with greater 
statistical accuracy, and have established the consistency between 
the two analyses. 

A fundamental assumption of the droplet theory on aging 
processes\cite{FH-88-NE} is that 
coarsening of domain walls, in case of spin glasses, is driven by
successive nucleation of thermally activated droplets. In order to
test this assumption, we have studied largest relaxation times
$\tau_{L}(T)$, which are those needed for a global spin-flip in each
sample with relatively small sizes $L$. We have found that the
$L$-dependence of $\tau_{L}(T)$ is consistent with the growth law of
domains, i.e., $R(t \sim \tau_{L}(T)) \sim L$. 
Furthermore, the temperature dependence follows in fact an Arrhenius 
law.

In our simulated results, however, there remain some discrepancies
with the original droplet theory. 
Firstly, it is assumed to grow as $L^\psi$
in the droplet theory, while the barrier free energy simulated grows
logarithmically with $L$. This results agree with the previous
conjectures extraxted from the $t$-dependence of 
$R(t)$.\cite{Kisker-96,Rieger93}. Secondly, the width of the
distribution of the barrier free energy does not increase with $L$. 
 
The present paper is organized as follows.
After describing the model and numerical method in the next section, 
we present and discuss the results of our simulation 
in \S~\ref{sec:fss-3}. Section~\ref{sec:fss-4} is devoted to the 
concluding remarks on the present work.

\section{Model and Numerical Method}
\label{sec:fss-2}

We carry on simulations on the 3D EA Ising spin-glass model with 
nearest-neighbor interactions $\{J_{ij}\}$ defined by the Hamiltonian,
\begin{equation}
        H = -\sum_{\langle ij \rangle}J_{ij}S_i S_j,
\label{eq:fss-2-1}
\end{equation}
on an $N=L\times L\times L$ simple cubic lattice.
The periodic boundary conditions are adopted.
The quenched random interaction $\{ J_{ij} \}$ are drawn from a Gaussian
distribution with zero mean and variance one. It has been reported that this
model has a spin-glass ordered phase below the critical temperature
$T_{\rm c}$ whose most recent  value is reported as 
$T_{\rm c}=0.95\pm 0.04$.\cite{Marinari-cd98-PS}
We use the single-spin-flip heat-bath Monte Carlo method.
As a unit time, which we call one Monte Carlo Step (MCS) hereafter, 
we adopt the time in which $N$ spins are updated.

The energy per spin $e_{T}(t)$ at $t$ MCS after the
temperature quench is calculated as
\begin{equation}
        e_{T}(t) = \frac{1}{N}\left[ \frac{1}{2t_{\rm ta}+1}
        \sum_{\tau=t-t_{\rm ta}}^{t+t_{\rm ta}}H(\tau) \right]_{\rm av},
\label{eq:fss-2-6}
\end{equation}
where $H(\tau)$ is a value of Hamiltonian at $\tau$ MCS. In order to
reduce thermal fluctuations we take here an average over short time 
$2t_{\rm ta}+1$ around $t$ with $t_{\rm ta}=t/1000$. 
The bracket $[\cdots]_{\rm av}$  
indicates the average over $N_{\rm s}$ Monte Carlo runs with
independent realizations of quenched random variables $\{J_{ij}\}$,
initial spin configurations, and random numbers. 
We have simulated systems with relatively
small sizes ($L=4 \sim 10$) in order to study the crossover between
the dynamic and static regimes as mentioned in \S 1, as well as 
large systems with $L=24$ and $32$ in order to 
obtain bulk behavior which is not affected by finite size effects
within our time window ($\sim 10^{6}$ MCS).

Following the previous works,\cite{Huse-91,Kisker-96,Marinari-98-VFDT}
the length scale of ordered domains $R(t)$ is estimated by analyzing 
the replica-overlap function, $G(r,t)$, defined as 
\begin{equation}
        G(r,t) = 
        \frac{1}{N} \sum_{i=1}^{N} \left[ 
        S_i^{(\alpha)}(t)S_i^{(\beta)}(t)
        S_{i+r}^{(\alpha)}(t)S_{i+r}^{(\beta)}(t) \right]_{\rm av}.
\label{eq:fss-2-2}
\end{equation}
Here $\alpha$ and $\beta$ are indices of the two replicas which have
different random spin configurations at $t=0$ (an instant of the 
temperature quench), and are updated independently.
Its correlation length is considered to be proportional to $R(t)$. 
In  eq.~(\ref{eq:fss-2-2}) $r$ is a spatial distance which we take 
only along the directions of the lattice axes. 
We have simulated large systems with $L=24$ and $32$ in this analysis.

\section{Results and Discussions}\label{sec:fss-3}

\subsection{Finite-size scaling of $e_{T}(t)$ }\label{subsec:fss-3-2}

\begin{figure}
\leavevmode\epsfxsize=85mm
\epsfbox{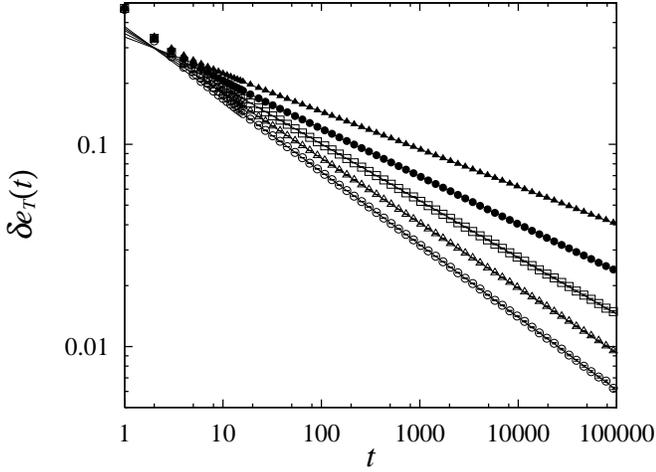}
\caption{The power law decay of excess energy per spin 
$\delta e_{T}(t)$ in the dynamic regime. The constants
$e_{T}^{(\infty)}$ are obtained by fitting the simulated date to 
eq.~(\ref{eq:fss-3}). The data are at 
$T=0.4, 0.5, 0.6, 0.7, 0.8$ from top to bottom. The systems simulated are
with $L=32$ and $N_{\rm s}=1600$ for $T=0.8$ and $0.7$, and with $L=24$ and 
$N_{\rm s}=3200$ for $T=0.6 \sim 0.4$. They are sufficiently large to obtain 
size-independent (bulk) energy decay.}
\label{fig:fss-3}
\end{figure}

Let us begin with analysis on the relaxation of the energy per spin
$e_{T}(t)$ of large systems with $L=32$ and $24$ after the quench. 
The log-log plots of $\delta e_{T}(t)\equiv e_{T}(t)-e_{T}^{(\infty)}$ 
vs. $t$ are shown in Fig.~\ref{fig:fss-3}, where $e_{T}^{(\infty)}$
denotes lim$_{t \rightarrow\infty}$lim$_{L \rightarrow \infty}e_T{}(t)$. 
We fitted the data to the following formula,\cite{Marinari-98-VFDT}
\begin{equation}
   \delta e_{T}(t) \equiv e_{T}(t) - e_{T}^{(\infty)} = 
   c t^{-\lambda{}(T)},
\label{eq:fss-3}
\end{equation}
with $e_{T}^{(\infty{})}$, $c$ and $\lambda$ being the fitting
parameters. From eq.~(\ref{eq:fss-1-1})  with
$R(t)\propto t^{1/z(T)}$ we obtain 
\begin{equation}
  \lambda = (d-\theta)/z.
\label{eq:fss-scaling-relation}
\end{equation}
As seen in Fig.~\ref{fig:fss-3}, the fitting works very 
well. The values of exponent $\lambda$ depend on temperature. 
It is proportional to $T$ at low temperatures
in agreement with the previous work\cite{Marinari-98-VFDT}.
Below  we determine $z(T)$ and check 
eq.~(\ref{eq:fss-scaling-relation}) numerically.

\begin{figure}
\leavevmode\epsfxsize=85mm
\epsfbox{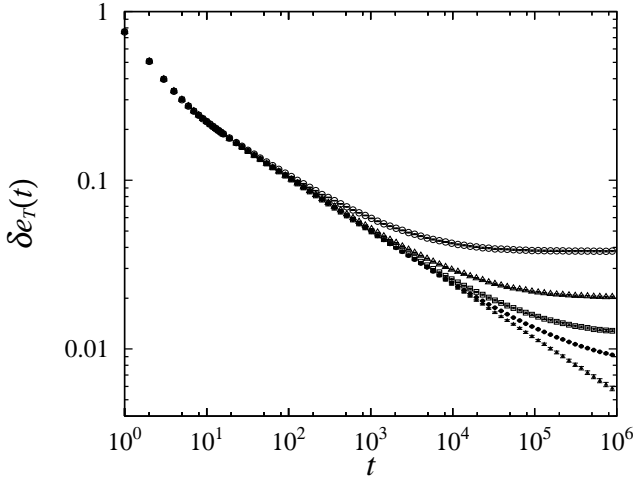}
\caption{
The decay of excess energy per spin $\delta e_T (t)$
for systems with $L=4$ (circles), $5$ (triangles), $6$ (squares), 
$7$ (solid circles), and $32$ (solid triangles). The asymptotic
energy per spin $e_{T}^{(\infty)}$ is determined as $-1.66204$ by fitting 
data of $L=32$ to eq.~(\ref{eq:fss-3}). The sample numbers $N_{\rm s}$ 
are $10^5$ for $L=4$, $5$, $6$ , $3.5\times 10^5$ for $L=7$, and $640$ for
$L=32$.
}
\label{fig:fss-4}
\end{figure}

In Fig.~\ref{fig:fss-4} we show the log-log plot of $\delta e_{T}(t)$ vs. 
$t$ at $T=0.7$ for small systems with $L=4,5,6,7$. 
It is seen that there exist two time regions of the relaxation curves.
In the shorter time region the curves have no size dependence (the
dynamic regime), while in the longer time region the curves depend on $L$
 and $e_{T}(t)$ approaches to an equilibrium value $e_{T}^{(L)}$ which 
also depends on $L$ (the static regime). 
The crossover between the two regimes is expected to occur when a typical 
size of ordered domain $R(t)$ reaches $L$. We therefore have tried some 
finite-size-scaling (FSS) analyses based on eq.~(\ref{eq:fss-tmp}). 
It has turned out that the following FSS ansatz works best:
\begin{eqnarray}
  \delta e_{T}(t) & = & 
     e_{T}(t)-e_{T}^{(\infty)}=L^{\theta-d}\tilde{f}(t/L^{z(T)}),
     \label{eq:fss-5}\\
  \tilde{f}(x) & = & \left\{
     \begin{array}{lll}
        c x^{-\lambda{}(T)} & \mbox{for} & x \ll 1,\\
        \Upsilon'(T) &\mbox{for} & x\gg 1,
     \end{array}\right.
     \label{eq:fss-6}
\end{eqnarray}
where $\Upsilon{}'{}(T)$ is a constant proportional to
$\Upsilon{}(T)$  in eq.~(\ref{eq:fss-1-1}).
In other words, the above FSS function has the following asymptotic limits:
\begin{equation}
        e_{T}(t) - e_{T}^{(\infty)} = \left\{ \begin{array}{ll}
                \Upsilon{}(T) R(t)^{\theta-d} & \mbox{for dynamic regime},\\
                \Upsilon'{}(T) L^{\theta-d} & \mbox{for static regime},
        \end{array}\right.
\label{eq:fss-3-2}
\end{equation}
with $R(t)$ being proportional to $t^{1/z(T)}$. 

In practice, we have performed the above FSS fit in the following way. 
By making use of $e_{T}^{(\infty)}$ evaluated above using the bulk data, 
and expecting that $\delta e_{T}(t)$ 
varies as a function of $x\equiv\log t$ and $y\equiv\log L$, we fitted
the data of $\delta e_{T}(t)$ to eq.~(\ref{eq:fss-5}) by means of 
the following trial function with ten parameters,
\begin{eqnarray}
        \log(e_{T}(t) - e_{T}^{(\infty)}) & = &
        (\theta-d) \log L + \log\tilde{f}(t/L^{z(T)})\\
        & \simeq & (\theta-d) y + \sum_{n=0}^{7}\frac{c_n}{n!}
        \left(x-z(T)y\right)^n.
\label{eq:fss-2-7}
\end{eqnarray}
Here we have expanded logarithm of $\tilde{f}$ with respect to logarithm of 
its argument up to the seventh order. The above fitting has been
done for 10 independent $e_{T}(t)$, each of which is obtained by
averaging over $N_{\rm s}/10$ samples. We have then estimated
errors of exponent $\theta$ and $z(T)$ from its variance over the 10
independent values.

\begin{figure}
\leavevmode\epsfxsize=90mm

\epsfbox{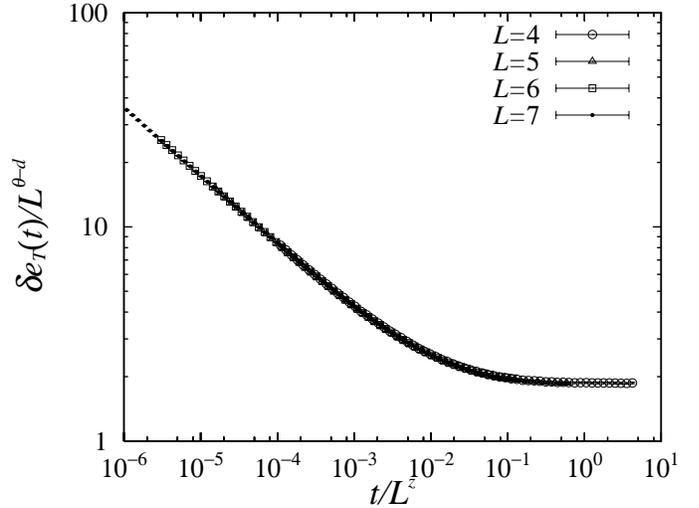} 
\caption{
The scaling plot of $\delta e_T(t)$ versus $t/L^{z(T)}$.}
\label{fig:fss-5}
\end{figure}
 
The results of the above FSS are shown in Fig.~\ref{fig:fss-5}.
We see that the scaling works quite satisfactorily. 
The similar results are obtained also at $T=0.8$, 
for which we have examined systems with $L=4,6,8,10$ and $N_{\rm s}$ 
up to $130000$. The exponents obtained are listed on 
Table~\ref{tbl:theta} for 
$\theta$ and Table~\ref{tbl:z} for $z(T)$. 
The values of $z(T)$ are consistent with those extracted from $R(t)$
in the larger systems, as will be discussed in the next subsection.
We therefore consider that the present FSS with eqs.~(\ref{eq:fss-5})
and~(\ref{eq:fss-6}) is quite reasonable. The exponent
$\theta$ seems almost independent on temperature as expected. 
Actually, its values derived by the present FSS at $T=0.7$ and $0.8$ 
are almost equal to the droplet exponent $\theta=0.19\pm 0.01$ 
extracted from the defect energy analysis at $T=0$ by Bray and 
Moore.\cite{BM-84} 
This result in the dynamic regime confirms the domain coarsening 
picture described in \S 1.

It is worth noting that static energy of finite size systems is 
higher than that of the infinite one by an amount $L^{\theta}$, 
that is, 
\begin{equation}
  \left({L \over l_0}\right)^{d}e_{T}(t) = 
 \left({L \over l_0}\right)^{d}e_{T}^{(\infty)} 
+ \Upsilon'(T) \left({L \over l_0}\right)^{\theta}.
\end{equation}
This is interpreted that the imposed periodic boundary conditions 
distort the spin correlations with respect to  what would 
be realized in an infinite system.\cite{Newman-98}
A similar and more transparent situation can be seen in a pure 
ferromagnet  with the {\it anti}-periodic boundary condition. 
The latter induces competition or frustration in the loop of 
the interactions winding the whole system.
Clearly it leaves a domain wall even in equilibrium. 
The FSS of eqs.~(\ref{eq:fss-5}) and~(\ref{eq:fss-6}) 
works well also for this case and the well known exponents, $z=1/2$
and $\theta=d-1$ are deduced. The situation in the present
spin-glass model is more complicated since we cannot separate the
effects of frustrations in bulk and of the constraint by the boundary 
conditions. However, the near agreement of the exponents $\theta$ at 
$T=0.7$ and $0.8$ with the one at $T=0$ suggests that the periodic 
boundary condition alone introduces a `defect energy' similar to the 
one at $T=0$ discussed by Bray and Moore. 
In this respect let us  note that a similar algebraic size 
dependence of finite size correction to the ground state energy with
the periodic boundary condition  
has been reported recently.\cite{Hartmann-97}

\subsection{Domain growth after quench}\label{subsec:fss-3-1}
Here we examine the replica-overlap function $G(r,t)$ of 
eq.~(\ref{eq:fss-2-2}) and the domain growth $R(t)$ after the quench.
Our analysis follows the work of Kisker et al.\cite{Kisker-96} 
We use an improved  method of evaluating $R(t)$ from $G(r,t)$ as
explained below. 
Because of the periodic boundary conditions, relation $G(r,t)=G(L-r,t)$
holds, and so $R(t)$ is evaluated as\cite{Cooper-82} 
\begin{equation}
        R(t) = 2\cdot\frac{L}{2\pi}
        \sqrt{\frac{\tilde{G}(0,t)}{\tilde{G}(2\pi/L,t)}-1},
\label{eq:fss-2-3}
\end{equation}
where
\begin{equation}
        \tilde{G}(k,t) = 2\int_{0}^{L/2}G(r,t)\cos{}kr \mbox{\rm d}r.
\label{eq:fss-2-4}
\end{equation}
If $G(r,t)$ is simply proportional to $\cosh[2(r-L/2)/R(t)]$
(or $\exp(-2r/R(t))$ in the limit $L \rightarrow \infty$), the
estimation of eq.~(\ref{eq:fss-2-3}) gives $R(t)$ exactly. We consider 
that our estimation of $R(t)$ is superior to the method
based on the moments $\int_0^{L/2}r^nG(r){\rm d}r$ with $n=0,1$ used
by Kisker et al. since our estimation significantly reduces the 
irrelevant contribution from small $r$ which deviate from an 
exponential law.
\begin{figure}
\leavevmode\epsfxsize=85mm
\epsfbox{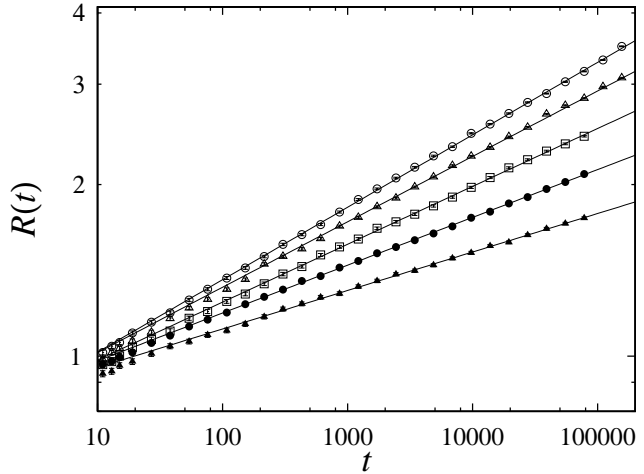}
\caption{
The length scale $R(t)$ of ordered domains 
estimated from
eq.~(\ref{eq:fss-2-3}) for temperatures $T=0.4, 0.5, 0.6, 0.7,$ and
$0.8$ from bottom to top. The solid lines are fitting lines $R(t) \propto
t^{1/z(T)}$.}
\label{fig:fss-yobun}
\end{figure}
\begin{figure}
\leavevmode\epsfxsize=85mm
\epsfbox{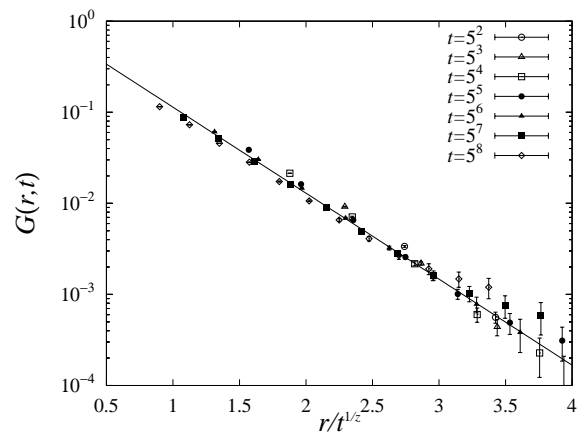}
\caption{
Logarithms of the replica-overlap function $G(r,t)$ versus 
$r/t^{1/z(T)}$ at $t=5^{n}$ ($n=2,\dots,8$) and $r>3$ for $T=0.7$
with $L=32$ and $N_{\rm s}=3500$.
The solid line is only a guide to the eye.}
\label{fig:fss-1}
\end{figure}

The averaged domain size $R(t)$ estimated from 
eqs.~(\ref{eq:fss-2-3})~and~(\ref{eq:fss-2-4}), and shown in 
Fig.~\ref{fig:fss-yobun}, are well fitted to 
$R(t)=b t^{1/z(T)}$. The obtained exponents $z(T)$ are 
$7.86(3)$, $8.71(3)$, $9.84(5)$, $11.76(6)$ and $14.80(7)$ at
$T=0.8$, $0.7$, $0.6$, $0.5$ and $0.4$, respectively.
These values at $T=0.8$, $0.7$ listed on Table~\ref{tbl:z}
are in good agreement with those
extracted from the FSS analysis on $\delta e_{T}(t)$.
\begin{table}
\caption{Exponent $\theta$ evaluated by our finite-size-scaling 
  (FSS) analysis on the energy relaxation. }
\label{tbl:theta}
\hspace*{\fill}
\begin{tabular}{ccc}\hline
$T$ & FSS & Bray and Moore\\ \hline
$0.8$ & 0.20(2) & --- \\
$0.7$ & 0.19(3) & --- \\
$0.0$ & --- & 0.19(1) \\ \hline
\end{tabular}\hspace{\fill}
\end{table}
\begin{table}
\caption{Exponent $z(T)$ estimated from our finite-size-scaling (FSS)
  analysis on the energy relaxation and directly from correlation
  length of $G(r,t)$. The values 
  are consistent with each others. The values $z(T)$ of the previous works 
  \cite{Marinari-98-VFDT,Kisker-96} are also
  listed, where the values of Marinari et al. are calculated from
  their estimation  as $1/z(T)\sim 0.16T$. 
  The present result is consistent with $z(T)$ by Marinari 
  et~al. but not with $z(T)$ by Kisker et al.}\label{tbl:z}
\begin{tabular}{cccc}\hline
$T$   & FSS & $R(t)$ & previous works \\ \hline
$0.8$ & $7.73(7)$ & $7.86(3)$ & $\sim 7.81$ (Marinari et al.)\\
$0.7$ & $8.83(8)$ & $8.71(3)$ & $\sim 8.93$ (Marinari et al.) \\ 
      &           &           &  $12.4$ (Kisker et al.)\\ \hline
\end{tabular}
\end{table}
The temperature dependence of $z(T)$ is given by $1/z(T) \simeq 0.17T$ at
$T<0.7$.  This linear $T$-dependence of $1/z(T)$ is consistent with 
the results by Marinari et al.,\cite{Marinari-98-VFDT} and those by
Kisker et al.\cite{Kisker-96}, though its coefficient $0.17$ is 
different from the one estimated from those of Kisker et al.

The replica-overlap function $G(r,t)$ is well scaled when it is plotted
against $r/t^{1/z(T)}$ as shown in Fig.~\ref{fig:fss-1}. Furthermore
it is seen that $G(r,t)$ in a longer length scale decays in a simple
exponential form. The exponents 
$z(T)$ used in this plot are those extracted by the fit of $R(t)$.

\subsection{Distribution of barrier free-energy of isolated droplets}
\label{subsec:fss-3-3}

As mentioned in \S 1, another important purpose of the present work
is to study directly relaxational dynamics of droplets. Since,
however, it is hard to specify each droplet in bulk, we examine here 
characteristic time scales $\tau_L(T)$ for spin configurations of 
systems with small $L$ to turn over as a whole in equilibrium. 
We regard such systems as isolated droplets of size $L$.

At first, $t_0$ MCS is elapsed starting from a random spin
configuration. Then the spin configuration $\{ S_i^{(\alpha)}(t_0) \}$
is copied to another replica $\beta$ and the two replicas are updated 
independently with different sequence of random numbers. The latter is
continued until  the clone-correlation function 
\begin{equation}
        Q(t)=\frac{1}{N}\sum_{i=1}^{N}\
        S_i^{(\alpha)}(t)S_i^{(\beta{})}(t).
\label{eq:fss-2-10}
\end{equation}
first crosses zero at some time $t_1$. Then the spin configuration  
$\{ S_i^{(\alpha)}(t_1) \}$ is copied to replica $\beta$ and the above 
process is repeated until $t_2$ when $Q(t)$ crosses zero. The
procedure is repeated once more, and the time when 
$Q(t)$ crosses zero is denoted by $t_3$.   

\begin{figure}
\leavevmode\epsfxsize=85mm
\epsfbox{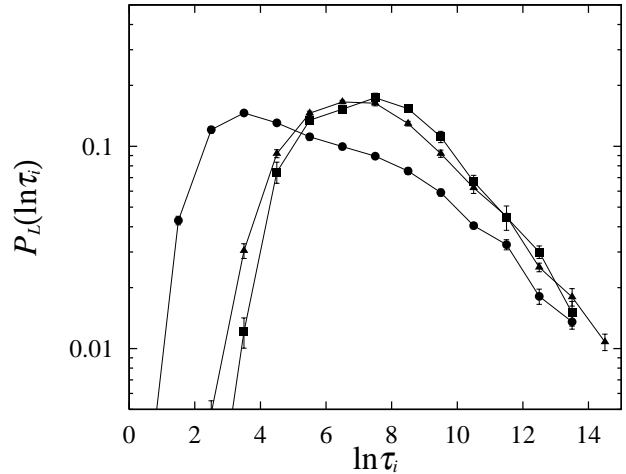} 
\caption{The distribution of global-flip times $\tau_1$ (solid
  circles), $\tau_2$ (solid triangles) and $\tau_3$ (solid squares), 
at $T=0.8$ in $L=4$ systems.
}
\label{fig:fss-6}
\end{figure}

In Fig.~\ref{fig:fss-6} we show distribution of the logarithmic times 
$P_{L}(\ln \tau_i)$, where $\tau_i=t_i - t_{i-1}$ with $t_0=0$. 
One can see clearly that $P_{L}(\ln \tau_2)$ and $P_{L}(\ln \tau_3)$
coincide with each other, but it is not the case for $P_{L}(\ln
\tau_1)$. The result is interpreted that $P_{L}(\ln \tau_1)$ is 
affected by off-equilibrium effects, while $P_{L}(\ln \tau_2)$ can be
regarded as an equilibrium distribution which we analyze in detail
below with $\tau_L(T) = \tau_2\ (=\tau)$. In this analysis we have
examined both the periodic and free boundary conditions. 

\begin{figure}
\leavevmode\epsfxsize=85mm
\epsfbox{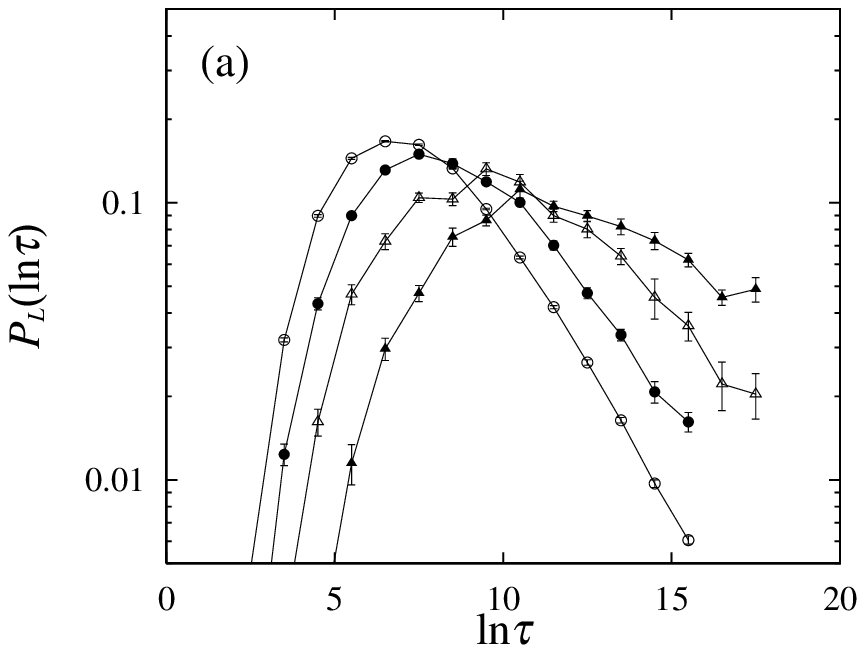}\\
\leavevmode\epsfxsize=85mm
\epsfbox{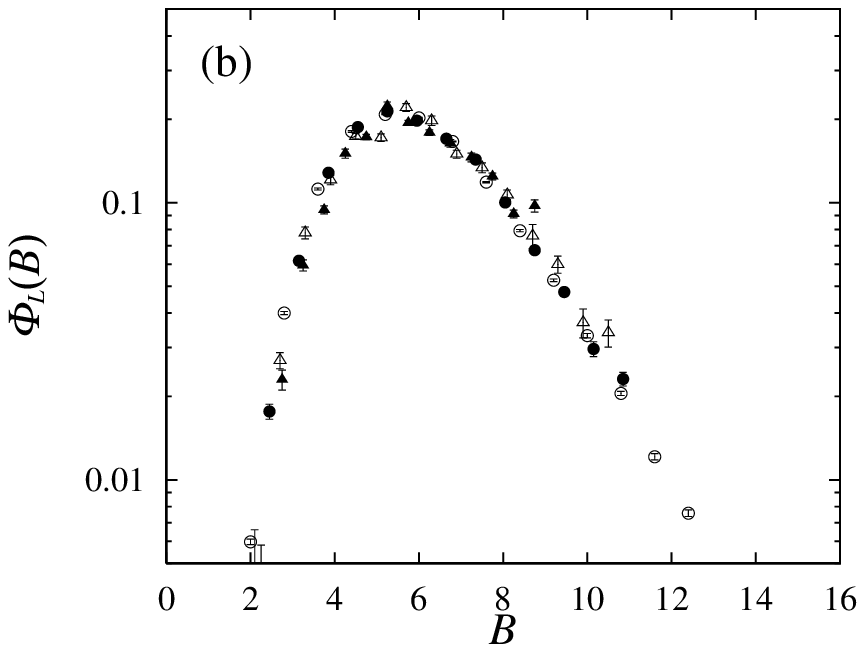}
\caption{(a) The distribution of $\tau\ (=\tau_L(T))$ for $L=4$ at 
$T=0.8$ (open circles), $0.7$ (solid circles),
$0.6$ (open triangles) and $0.5$ (solid triangles).
(b) The plot of $\Phi_{L}(B)$ extracted from eq.~(\ref{eq:fss-2-9}).
}
\label{fig:fss-7}
\end{figure}

In Fig.~\ref{fig:fss-7}-(a) we show the log-log plot of $P_{L}(\ln
\tau )$ versus $\tau$ of the system with $L=4$ at $T=0.8$, $0.7$,
$0.6$ and $0.5$. If the relaxation process 
is of the Arrhenius-type, i.e., $\tau = \tau_0 \exp (B/T)$, 
the distribution of the free-energy barriers $\Phi_{L}(B)$ is
written as 
\begin{equation}
        \Phi_{L}(B) = \frac{1}{T}P_{L}(\ln\tau{}/\tau_{0})
\label{eq:fss-2-9}
\end{equation}
where $\tau_{0}^{-1}$ is the attempt frequency. 
With  $\tau_0=1$ the data in Fig.~\ref{fig:fss-7}-(a) lie top on each
other as shown in Fig.~\ref{fig:fss-7}-(b). This result indicates that 
the relaxational dynamics of present interest is in fact an
Arrhenius type whose barrier free energy $B$ little depends on 
temperature.

\begin{figure}
\leavevmode\epsfxsize=85mm
\epsfbox{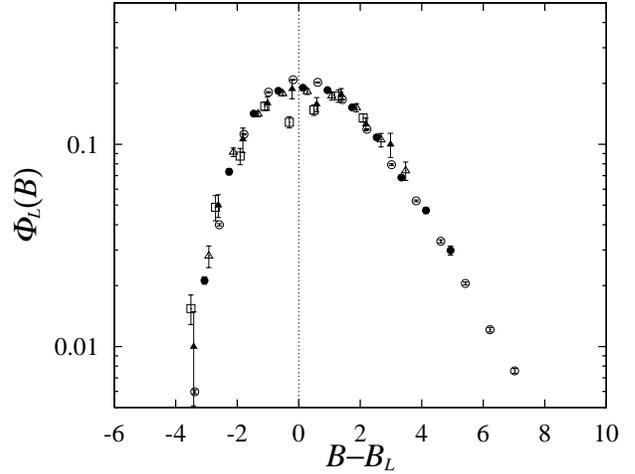}
\caption{
The plot of $\Phi_{L}(B)$ versus $B-B_{L}$, at $T=0.8$ for 
$L=4$ (open circles), $6$ (solid circles), $8$
(open triangles), $10$ (solid triangles) and $12$ (open squares).
}
\label{fig:fss-8}
\end{figure}

Making use of relation $B=T\ln \tau$ confirmed above, we next examine
the $L$-dependence of $\Phi_{L}(B)$. In Fig.~\ref{fig:fss-8} we
show the plot of $\Phi_{L}(B)$ against $B-B_L$ at $T=0.8$, where 
$B_L$ is the value, at which $\Phi_{L}(B)$ is maximum for each
$L$. The resultant $B_L$ is well fitted to 
\begin{equation}
        B_{L} = \Delta' \ln L,
                \label{eq:fss-3-5}
\end{equation}
as shown in Fig.~\ref{fig:fss-9} below. 
Another interesting observation in Fig.~\ref{fig:fss-8} is that the
shape of the distribution $\Phi_{L}(B)$ does not depend on
$L$, while its peak position does. 
This is certainly in disagreement with the original droplet
theory,\cite{FH-88-EQ} in which the width of $\Phi_{L}(B)$ is
conjectured to be of the same order as that of its mean.

\begin{figure}
\leavevmode\epsfxsize=85mm
\epsfbox{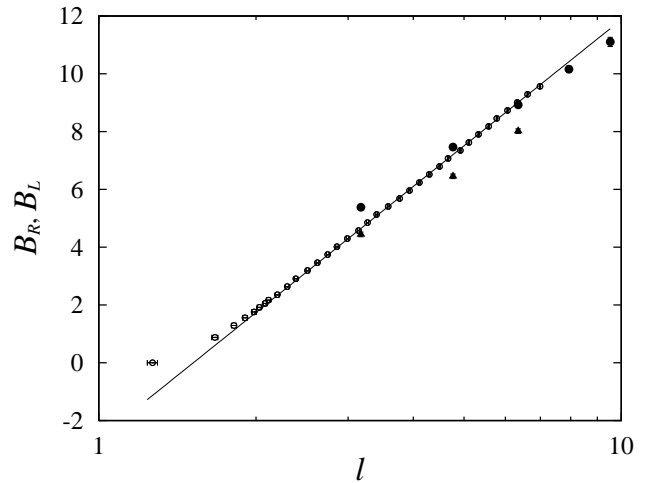}
\caption{The characteristic barrier free energy corresponding to 
characteristic length scale $l$ (at $T=0.8$). 
The open circles are estimated from the growth of $R(t)$.
The solid circles and solid triangles are estimated from the distribution
of global flip times $\tau_L(T)$ with the periodic and open boundary
conditions, respectively.
}
\label{fig:fss-9}
\end{figure}

If the coarsening process of domains represented by $R(t)$ analyzed
before is also assumed due to nucleation of domains of the
corresponding size by thermally activated process, the associated
barrier free energy $B_R$ is given by\cite{Kisker-96} 
\begin{equation}
        B_{R} = \Delta \ln R.
                \label{eq:fss-3-8}
\end{equation}
In Fig.~\ref{fig:fss-9} we plot $B_{R}$ thus evaluated
and  $B_{L}$ of eq.~(\ref{eq:fss-3-5}) against the properly scaled
length $l$. For $B_{R}$ we set $l=2R(t)$, while for $B_{L}$
we use $l=L/(2^{1/d})$ since the clone-correlation function   
can cross zero when a half of spins in the system turns over. 
The logarithmic dependence on $l$ of both $B_{R}$ and $B_{L}$ is
clearly seen in the figure. 
Also it is seen that the coefficient $\Delta$ for $B_{R}$
is a little larger than $\Delta'$ for $B_{L}$. This difference may 
indicate that dynamics of ordered domains in a large system is not
perfectly independent from their neighboring ones as was assumed
in the original droplet theory. 
A similar conclusion has been also 
deduced experimentally.\cite{Weissman}  

\subsection{Discussions}
Here let us comment on the magnitude of energy, 
length and time scales in the aging process simulated so far.
Our result is consistent with $F_{L}\sim L^{\theta}$, while the
barrier free energy $B_L$ associated with them is proportional to 
$\ln L$. It  
is then naturally expected that the latter is overcome by the former at a
certain value of $L$, and so that the present results cannot provide us a 
proper picture in the thermodynamic limit. 
But we note that $F_{L}$ evaluated from the present simulation is given by 
\begin{equation}
        F_{L} \propto N\delta e_{T} \simeq 2.0 \times (L/l_0)^{\theta},
\end{equation}
with $\theta = 0.19$, where the factor $2.0$ is determined from the
saturated value in the scaling plot of Fig.~\ref{fig:fss-5}.
Because of this small value of $\theta$, combined with a relatively
large value of $\Delta\ (\simeq \Delta')$ in eq.~(\ref{eq:fss-3-8})
which in turn is due to a relatively large value of $z(T)(\simeq 8
\sim 10)$, $F_{L}$ and $B_R$ (or $B_L$) become of the same order of
magnitude only when $L$ is of the order of $10^8\sim 10^9$ with $l_0=1$.
The corresponding time scale, for $R(t)$ to reach to this length
scale, is much more than an astronomical one. In this context, it is
pointed out that $B_R$ experimentally extracted by Joh et al.\cite{Joh-cd98}
recently is consistent with the simulated results expressed by
eq.~(\ref{eq:fss-3-8}) (with $\tau/\tau_0 \sim 
10^{12}$ and $R/l_0 \sim \ {\rm sevral}\ 10$). We therefore consider
that the results obtained in the present work are applicable to
aging phenomena observed within a time window of experiments on realistic
spin glasses. However, mechanism which gives rise to the ln$R$
dependence of $B_R$ is not clear yet. It might be related to 
fluctuations governed by the $T=T_{\rm c}$ criticality since
temperatures investigated in the present work may not be far enough from 
$T_{\rm c}$. This problem is left for future work.

\section{Concluding Remarks}\label{sec:fss-4}

By means of Monte Carlo simulations we have studied aging dynamics
after the temperature quench from $T=\infty$ to $T$ in the spin-glass 
phase in the 3D Ising EA model. 
The results of the energy evolution $e_{T}(t)$, 
in particular, the finite-size scaling on $\delta e_{T}(t)\ (= e_{T}(t)
- e_{T}^{(\infty)})$, strongly suggest that the droplet picture
describes aging dynamics appropriately. 
It is emphasized that these results are obtained based on
eq.~(\ref{eq:fss-1-1}) which is a most fundamental ingredient 
of the droplet picture, and that they are extracted by
analyzing time evolution of one spin configuration as compared with
the previous works which were making use of the replica-overlap. 

Also by analyzing relaxation times $\tau_L$ required for a global 
flip of spin configurations in small systems (isolated droplets) of size 
$L$, we have shown that the $L$-dependence of $\tau_L$ is
consistent with the $t$-dependence of the mean domain size $R(t)$,
i.e.,  $R(t \sim \tau_L(T) ) \sim L$. 
This result further supports the droplet picture that coarsening of
domain walls is in fact driven by successive nucleation of thermally
activated droplets.
The explicit form of this $t\ (\tau)$-dependence of $R(t)\ (L)$, 
however, disagrees with the original droplet theory due to Fisher and
Huse:\cite{FH-88-NE,FH-88-EQ} 
the barrier free energy associated with ordered domains of length
scale $L$ is  proportional not algebraically as predicted by the their
theory but logarithmically to $L$.

\section*{Acknowledgments}
We would like to thank P. Nordblad, R. Orbach and E. Vincent for
discussions on their experimental works.  
Two of the present authors (T. K. and H. Y.) were supported by Fellowships 
of Japan Society for the Promotion of Science for Japanese Junior Scientists.
This work is supported by a Grant-in-Aid for International Scientific
Research Program, ``Statistical Physics of Fluctuations in Glassy
Systems'' (\#10044064) 
and by a Grant-in-Aid for Scientific Research Program (\#10640362),
from the Ministry of Education, Science and Culture.
The present simulation has been performed on FACOM VPP-500/40 at the 
Supercomputer Center, Institute for Solid State Physics, the University of 
Tokyo and on HITACHI SR2201 at the Computer Centre of the University of Tokyo.

\bibliography{bibcluster}

\end{document}